\documentclass[structabstract]{aa}  
\usepackage{graphicx}
\usepackage{lscape}
\usepackage{tabularx}
\usepackage{txfonts}
\usepackage{natbib}
\usepackage{sidecap}
\usepackage{comment}
\usepackage{color}
\usepackage{booktabs}

\begin{document}
\newcolumntype{C}[1]{>{\centering\arraybackslash}p{#1}}

\title{Identifying two groups of massive stars aligned in the $l\sim38^{\circ}$ Galactic direction}
\author{S. Ram\'irez Alegr\'ia\inst{1} \and A. Herrero\inst{2,3} \and K. R\"ubke\inst{2,3} \and A. Mar\'in-Franch\inst{4,5} \and M. Garc\'ia\inst{6} \and J. Borissova\inst{7,8}}

\institute{Unidad de Astronom\'ia, Universidad de Antofagasta, Antofagasta, Chile. \email{sebastian.ramirez@uamail.cl} 
\and Instituto de Astrof\'isica de Canarias, c/V\'ia L\'actea s/n, 38205 La Laguna, Tenerife, Spain. \and Departamento de Astrof\'isica, Universidad de 
La Laguna, E-38205 La Laguna, Tenerife, Spain. \and Centro de Estudios de F\'isica del Cosmos de Arag\'on (CEFCA), E-44001, Teruel, Spain. \email{amarin@cefca.es} 
\and Departamento de Astrof\'isica, Universidad Complutense de Madrid, E-38040, Madrid, Spain. \and Centro de Astrobiolog\'ia, CSIC-INTA. Ctra. Torrej\'on a Ajalvir km.4, 
E-28850 Torrej\'on de Ardoz, Madrid, Spain. \and Instituto de F\'isica y Astronom\'ia, Facultad de Ciencias, Universidad de Valpara\'iso, Valpara\'iso, Chile. \and Millennium 
Institute of Astrophysics, MAS.  \\}
\date{Received 7 August 2017 / Accepted ??}

\abstract
{Recent near-infrared data have contributed to unveil massive and obscured stellar populations in both new and previously
 known clusters in our Galaxy. These discoveries lead us to view the Milky Way as an active star-forming machine.}
{We look for young massive cluster candidates as over-densities of OB-type stars. 
The first search, focused on the Galactic direction $l=38^{\circ}$, resulted in the detection of two objects with a remarkable
population of OB-type star candidates.}
{With a modified version of the friends-of-friends algorithm AUTOPOP and using 2MASS and UKIDSS-GPS near-infrared ($J$, $H$, and $K$) 
photometry for one of our cluster candidates (named Masgomas-6) we selected 30 stars for multi-object and long-slit $H$- and $K$-spectroscopy. 
With the spectral classification and the near-infrared photometric data, we derive individual distance, extinction and radial velocity.}
{Of the 30 spectroscopically observed stars, 20 are classified as massive stars, including OB-types (dwarfs, giants and supergiants),
two red supergiants, two Wolf-Rayet (WR122-11 and the new WR122-16), and one transitional object (the LBV candidate IRAS 18576+0341).
 The individual distances and radial velocities do not agree with a single cluster, indicating that we are observing two populations of massive stars
in the same line-of-sight: Masgomas-6a and Masgomas-6b. The first group of massive stars, located at 3.9$^{+0.4}_{-0.3}$ kpc, contains   
both Wolf-Rayets and most of the OB-dwarfs, and Masgomas-6b, at $9.6\pm0.4$ kpc, hosts the LBV candidate and an evolved population
of supergiants. We are able to identify massive stars at two Galactic arms, but we can not clearly identify whether these massive stars
form clusters or associations.}
{}
 \titlerunning{Two groups of massive stars aligned in $l\sim38^{\circ}$.}
\maketitle


\section{Introduction}

In the last decade, the study and discovery of young and massive star clusters have received a vital boost, thanks to near-infrared 
all-sky and large-scale photometric surveys. These surveys have revealed regions of the Milky Way obscured by intense and variable 
interstellar extinction, unveiling the massive stars contained in deeply embedded clusters. The catalogues of cluster candidates 
detected using the near infrared surveys 2MASS \citep{skrutskie06}, GLIMPSE \citep{benjamin03}, UKIDSS-GPS \citep{lucas08}, and 
Vista-VVV \citep{minnitiVVV10,saito12} have reported close to 2000 objects. These catalogues have helped to improve the census of 
massive clusters (objects with total mass $M_T > 10^4 M_{\odot}$), but it remains incomplete; according to \citet{hansonpopescu08} 
and \citet{ivanov10} more than 100 massive clusters are expected to be part of the Milky Way. Less than 20 massive clusters have been 
reported in our Galaxy.
 
 Cluster candidates catalogues are mostly based on the detection of stellar over-densities and are biased towards certain kind of objects. 
 The recent results derived from the VVV survey include catalogues focused on very young populations (less than 5 Myr), with their 
  main-sequence stars still deeply embedded \citep{barba15}, less extinct objects characterized by a stellar over-density
 detected by eye \citep{borissova11,borissova14}, or by automated detection procedures \citep{solin14,ivanov17}. The confirmation of 
 candidates as true objects is possible after spectroscopic or astrometric follow-up. In the case of the \citet{borissova11} catalogue, the spectroscopic
 data revealed clusters with Wolf-Rayet stars \citep{chene13,chene15}, OB-type dwarfs \citep{ramirezalegria14a,ramirezalegria16}  or young
 stellar objects \citep{borissova16}, as part of their stellar population.

The MAssive Stars in Galactic Obscured MAssive clusterS (MASGOMAS) is an initiative to look for massive clusters by identifying 
over-densities of OB-star candidates as young stellar cluster candidates in the 2MASS and other IR catalogues.
In the initial stages of the project, using a set of near-infrared photometric cuts ($K_S < 12.0$, $(J-K_S)>1.0$ and the reddening-free parameter 
 $Q_{IR}$ between -0.2 and 0.2; see \citealt{comeron05, negueruela07}, and Section \ref{discovery}), we discovered one massive cluster in the direction of the close end
 of the Galactic Bar (Masgomas-1; \citealt{ramirezalegria12}) and one cluster with two cores of massive star formation (Masgomas-4; \citealt{ramirezalegria14b}). 
 
 As photometric cuts select OB-type stellar candidates with some level of reddening (limiting to $(J-K_S)>1.0$ to avoid foreground contamination), 
 MASGOMAS clusters are young (less than 10 Myr) and not very embedded objects, compared for example with the candidates from the 
 \citeauthor{barba15} catalogue. The stellar population in the MASGOMAS clusters is principally found in the main-sequence. It may be affected by differential 
 reddening but can be distinguished from the field stellar population by comparing the cluster and control photometric diagrams (colour-magnitude, colour-colour 
 and $Q_{IR}$-magnitude).
 
 As we explore deeper using near-infrared data, crowding and chance alignment are factors to be considered in the search of cluster 
 candidates. Kinematic information, complemented with the surveys near-infrared photometry and spectroscopic follow-up for
 the candidates' most probable star members are useful tools to separate the entangled stellar population of the aligned clusters and
 to identify the asterism in the catalogues.

In this article we report the discovery of a massive star population around two massive stars: the Wolf-Rayet 
\object{WR122-11} (a.k.a. WR1583-B73; \citealt{faherty14}) and the blue supergiant/LBV candidate \object{IRAS 18576+0341} 
\citep{ueta01,pasquali02}. We present the first candidates derived with the MASGOMAS automatic searching software in Section 
\ref{discovery}, the near-infrared spectrophotometric data in Section \ref{NIRdata}, followed by the spectral classification and the analysis 
of the massive star populations confirmed by the spectroscopy (Sections \ref{spec} and \ref{analysis}).

\section{Candidate discovery}\label{discovery}
The MASGOMAS project focuses on the search of over-densities formed by OB-type star candidates and the characterization of
these candidates using near-infrared photometry and spectroscopy. We select the OB-type candidates using three photometric
criteria: $K_S$ magnitudes less than 12 mag (a limit adopted by the instrument used in our spectroscopic follow-up), 
red $(J-K_S) > 1.0$ colours, and reddening-free parameter $Q_{IR}$ between -0.2 and 0.2 \citep{ramirezalegria12}. The
adopted $Q_{IR}$ parameter is:
\begin{equation}
Q_{IR} = (J-H) - 1.70\cdot(H-K_S)
\end{equation}
the Rieke extinction law \citep{rieke89}, with $R=3.09$ \citep{rieke85}. \footnote{The reddening-free parameter, is 
defined as $Q_{IR} = (J-H) - \frac{E(J-H)}{E(H-K_S)}\cdot(H-K_S)$,and the numerical value of $\frac{E(J-H)}{E(H-K_S)}$ depends on the extinction law.}

Limiting $Q_{IR}$ between -0.2 and 0.2, we clean the photometry from disc giant stars and focus our search
on OB-type dwarfs (with $Q_{IR}\sim0$). Main expected contaminants from our method are A and early-F dwarfs, although photometric errors
or extinction law variations may shift some late-type stars into this zone. The photometric diagrams for the massive cluster Masgomas-1 
\citep{ramirezalegria12} and the double-core young cluster Masgomas-4 \citep{ramirezalegria14b} support the use of this law to describe 
the extinction in this Galactic direction. After this first stage, when we detected cluster candidates as over-densities by eye, we 
developed an automatic searching software based on AUTOPOP \citep{garcia09,garcia11} and adapted to use 2MASS photometry.

\begin{table*}
\caption{Summary of Masgomas objects detected during different project phases.}
\begin{center}
\begin{tabular}{ccccC{10cm}}
\toprule
ID & l & b   & Detection method & Comments \\
     &  [deg]          & [deg]              &                              &  \\
 \midrule
 01 & 33.109 & +0.424  &  Visual detection   & Massive ($M>10^4 M_{\odot}$) cluster with confirmed OB and RSG population \citep{ramirezalegria12}. \\
 02 & 28.552 & +4.004  &  Visual detection   & Giant field stars asterism mimicking RSG cluster.   \\
 03 & 29.269 & +0.014  &  Visual detection   & Part of Alicante 10 \citep{gonzalez-fernandez12}.   \\
 04 & 40.530 & +2.577  &  Visual detection   & Two-cored cluster with confirmed young and massive population \citep{ramirezalegria14b}. \\
 05 & 40.362 &  -0.701  &  Automatic            & Known cluster \object{Juchert 3}, a.k.a. \object{DSH J1907.5+0617} \citep{kronberger06}. \\
 06 & 37.290 &  -0.221  &  Automatic            & $N_{mem}$=12, radius=62\arcsec,  $\sigma$= 0.0021 [$N_{mem}$/arcsec$^2$].  \\

\bottomrule
\end{tabular}
\end{center}
\label{autopop}
\end{table*}

 The original AUTOPOP algorithm consists of two main routines: an automatic finding of stellar groups and an isochrone analysis 
 over those groups. Using the photometric criteria to select OB-type candidates, we used the first routine to 
 detect over-densities of OB-type candidates. This routine, based on a friends-of-friends method, considers two stars as part of a
 group if their separation is less than a search distance $D_S$. To be considered as a cluster candidate, the number of members in 
 the group, $N_{mem}$, must be higher than a threshold $N_{min}$. For the test version of the algorithm we arbitrarily
 set the values of $D_S=60\arcsec$ and $N_{min}=10$. The first test was focused on a box region of $3 \degr \times 3 \degr$ and centred around $l=38^\circ$ 
 and $b=0^\circ$. This search reported two candidates with an interesting number of member candidates: Masgomas-5
 ($N_{mem}=15$; a.k.a. \object{Juchert 3}, \citealt{kronberger06}), and the new candidate selected for spectroscopic follow-up Masgomas-6 
 ($N_{mem}=12$). The cluster candidate parameters derived from the automatic algorithm are given in Table \ref{autopop}.
 
\begin{figure*}
\centering
\includegraphics[width=18cm,angle=0]{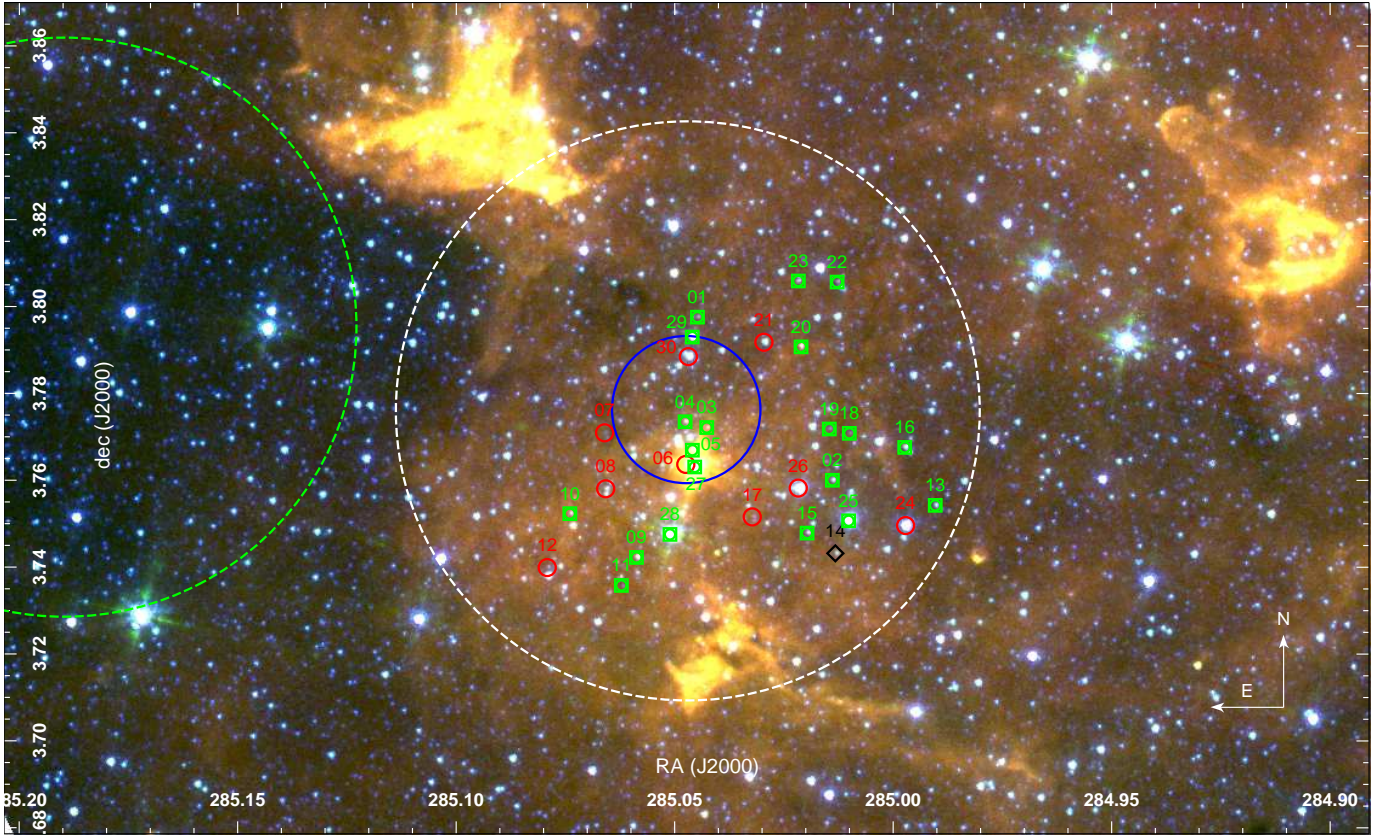} 
            \caption{GLIMPSE false colour (blue=IRAC I1, green=IRAC I3, red=IRAC I4) image for Masgomas-6. The extension of the cluster 
            candidate derived from the automatic search algorithm is shown with the solid blue circle. The areas used for decontamination are shown with 
            segmented white (cluster field) and green (control field) circles. We marked with small green boxes the stars spectrally classified as 
            early-type (dwarf, giant or supergiant) and the evolved massive stars (Wolf-Rayet); with red circles, we marked the late-type giants, 
            and with a black square the star without classification (star 14). The length of the orientation arrows is 1.0 arcmin. }
       \label{m06_spitzer}
\end{figure*}

 In Figure \ref{m06_spitzer} we present Masgomas-6 in a false colour GLIMPSE mid-infrared image. In the mid-infrared, it is 
 easy to observe an extended nebulosity surrounding the candidate. The extension of this nebulosity is larger than the
estimated radius for Masgomas-6, and as we demonstrate later, with the information derived from the observed spectra, the 
distribution of the stellar population goes beyond the first estimation of the cluster candidate radius.

After the discovery of the cluster candidate, we review the literature and found two objects, inside an area of 4$\arcmin$ around 
the center of Masgomas-6, that can spot the presence of massive stars: the Wolf-Rayet \object{WR122-11} \citep{faherty14} 
and the blue supergiant/LBV candidate \object{IRAS 18576+0341} \citep{ueta01,pasquali02}. There are no previous reports of a cluster 
hosting these two massive objects. 


\section{Observational data}\label{NIRdata}
 
 After detecting Masgomas-6 as a cluster candidate using the 2MASS photometry, we completed a near-infrared spectroscopic 
follow-up for the brightest most probable stellar members with LIRIS (at the William Herschel Telescope, Roque de Los Muchachos 
Observatory, La Palma; \citealt{manchado04}). For the selection of spectroscopic candidates and the photometric analysis of the 
candidate we used the photometry from the UKIDSS Galactic Plane Survey (GPS; \citealt{lucas08}). 

\subsection{Photometry}
The UKIDSS GPS is a near-infrared public legacy survey, completed with the Wide Field Camera (WFCAM) on the United
Kingdom Infrared Telescope (UKIRT). The survey covered the regions between: $141 < l < 230$, $-5 < b < 5$; $15 < l < 107$, 
$-5 < b < 5$; and $-2 < l < 15$, $-2 < b < 2$. The survey includes observations in $J$, $H$ and $K$ bandpasses, this last 
one observed in two different epochs. A detailed description of the UKIDSS GPS data and its attributes (or parameters, 
including photometric flags), available through the WFCAM Science Archive (WSA) webpage is given by \citet{lucas08}.
 
The UKIDSS GPS survey includes two flags particularly helpful to describe data quality: $mergedClass$ and $ppErrBits$. The 
first flag helps to discriminate elongated from rounded sources. Using only sources with $mergedClass = -1$ or $-2$, we include
only photometry from objects with probability of being stars larger than 70\%\footnote{http://wsa.roe.ac.uk/www/glossns\_p.html}.
For the $ppErrBits$ flag, which quantifies how reliable the photometry is, 98\% of the sources used in our analysis 
 have the highest quality flag (i.e., $ppErrBits < 256$, see http://wsa.roe.ac.uk/ppErrBits.html).
  
    Our saturation limits were $J=13.50$, $H=12.00$, and $K=11.00$ mag. These limits are slightly larger than suggested in the main 
  GPS paper ($H=12.25$, and $K=11.50$ mag). We observe a good behaviour of the sources in these limits after inspection of 
  the associated error and the $ppErrBits$ flags. We replaced the individual saturated photometry with the corresponding 2MASS 
  photometry, transformed to the UKIDSS bandpass system using the photometric transformation given by \citet{lucas08}, and 
  assuming a standard value of $E(B-V)=3$.
  
\begin{figure*}
\centering
\includegraphics[width=3.6in,angle=0]{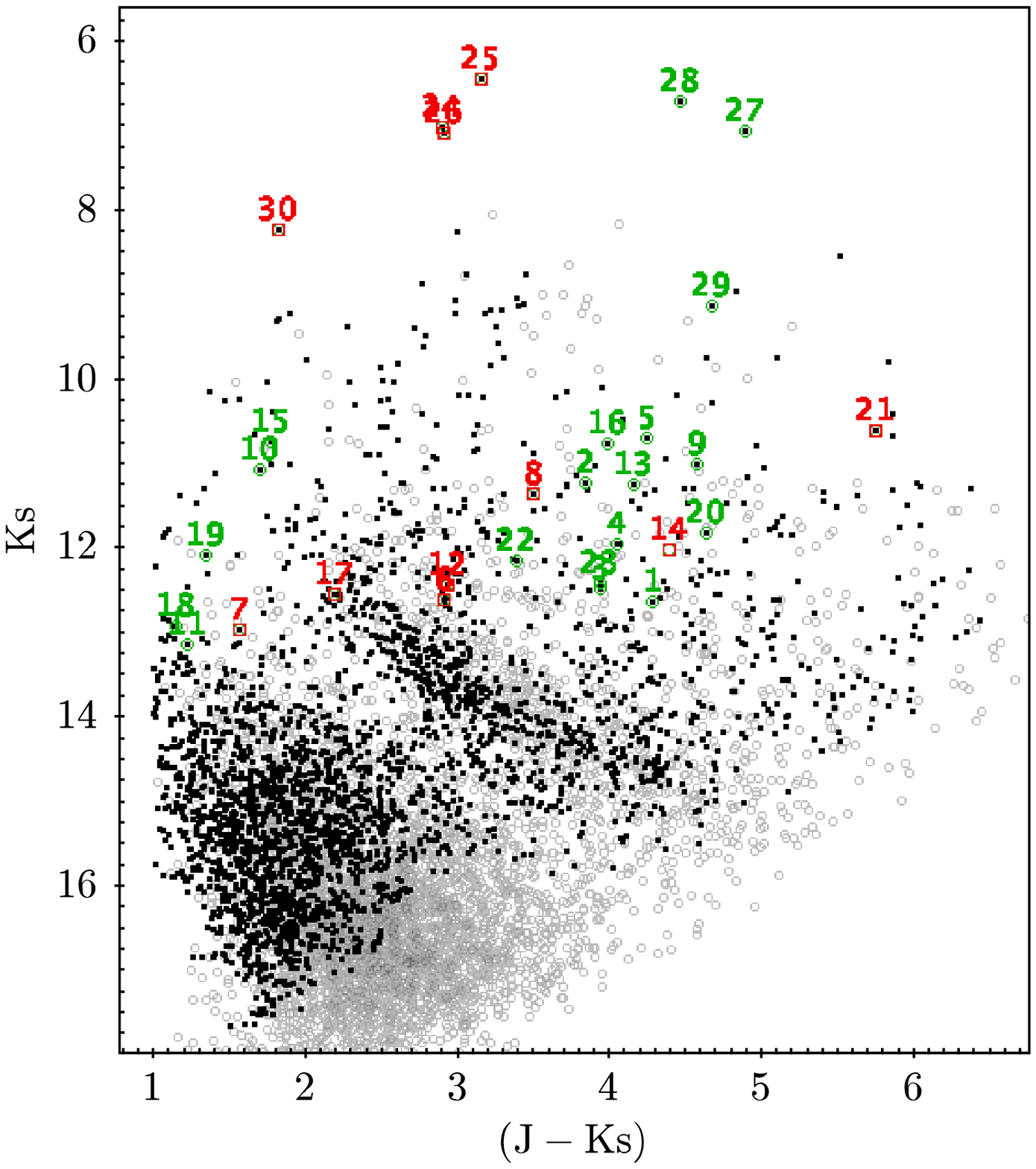} 
\includegraphics[width=3.6in,angle=0]{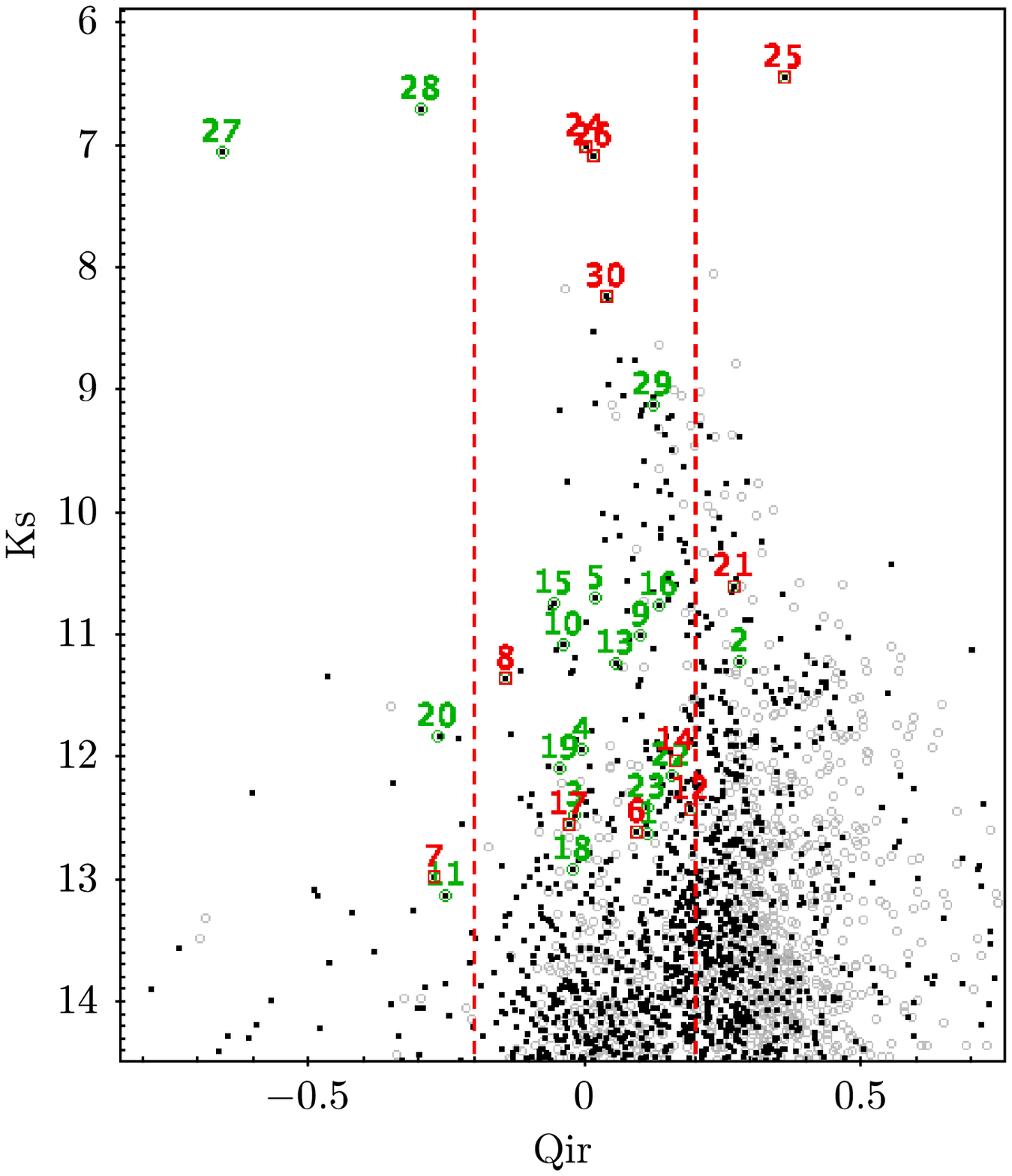} 
            \caption{Colour-magnitude (left) and free-reddening parameter--magnitude  $Q_{IR}$ (right) diagrams for Masgomas-6. Black dots show 
            the position of the most probable cluster members, and grey symbols show the position of the field stars. We include in the figure
            the spectroscopically confirmed massive stars with green open squares and the late-type giant stars with red open circles. The 
            individual identification numbers (ID, table \ref{data_stars}) are also shown on top of each colour symbol. In the pseudocolour-magnitude
            diagram, we can see that most of the massive stars are found between $Q_{IR}=-0.2$ and 0.2, as expected by our selection
            method.}
       \label{m06_cmd}
\end{figure*}

\begin{figure}
\centering
\includegraphics[width=3.8in,angle=0]{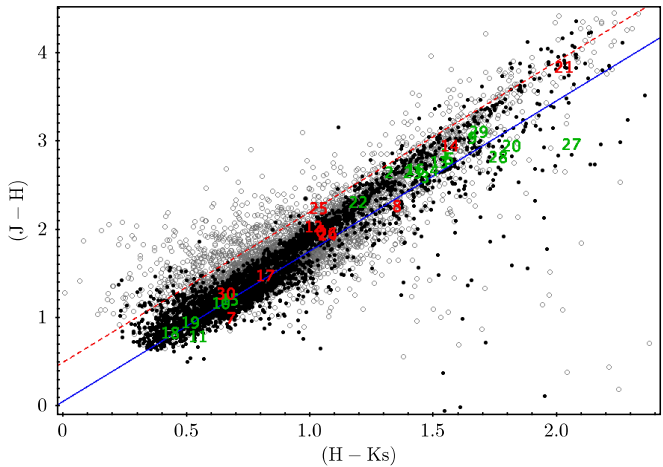} 
            \caption{Colour-colour diagram for Masgomas-6. Black dots show the position of the most probable cluster members, and grey symbols
            show the position of the field stars. Objects with spectroscopic data are shown with green (massive stars) and red symbols (late-type giants).
            In the figure we included the projection, following the extinction law by \citet{rieke89} (with $R=3.09$; \citealt{rieke85}), for an O giant 
            (blue continuous line; intrinsic colour by \citealt{martins06}) and a G5 giant (red segmented line; intrinsic colour by \citealt{cox00}). The stellar
            population located above the red-segmented line corresponds mostly to the disc giant population and is observed only in the grey-field
            stars. The massive star population, confirmed spectroscopically, is observed close to the continuous blue line and below the giant sequence.} 
       \label{m06_ccd}
\end{figure}

  Taking advantage of the better spatial resolution and depth, compared with the 2MASS photometry, we selected OB-type candidates 
 using UKIDSS GPS photometry and the same strategy of photometric criteria used to detect the cluster candidates. In Figures
 \ref{m06_cmd} and \ref{m06_ccd} we show the colour-magnitude, free-reddening parameter--magnitude ($Q_{IR}-K$) and colour-colour diagrams
 for Masgomas-6. The stars selected for spectroscopic follow-up are marked with red and green numbers, using the identification numbers
 from Table \ref{data_stars}. In the diagrams, we present the photometric information from the cluster and control fields shown in Figure \ref{m06_spitzer}. 
 In the $Q_{IR}-K_S$ diagram we see a vertical sequence of stars centered around $Q_{IR}=0$, not present in the control field. We interpret
 these object as candidates for the cluster OB-type population. 
 
 \subsection{Spectroscopy}
 
 For the spectroscopic follow-up, we use the LIRIS instrument, a near-infrared spectrograph (and imager) mounted 
at the Cassegrain focus of the 4.2 m William Herschel Telescope (La Palma, Canary Island, Spain). The camera is equipped with a
 Hawaii 1024$\times$1024 HgCdTe array detector, with a field of view of $4.2'\times4.2'$ and a spatial scale of $0\,\farcs25\mathrm{~pixel}^{-1}$.
 The instrument has multi-object (MOS) and long-slit (LS) spectroscopic modes and if the $H$ and $K$ pseudogrism are used, the resolution is
 $R\sim2500$ \citep{fragoso2008}. The near-infrared spectra were obtained on 2013 June 29 and 30, using both spectroscopic 
 observing modes.
 
 For the MOS we designed two masks, focused mainly on OB-type stellar candidates. Mask A contained twelve stars and mask B 
 thirteen stars. We selected each mask candidate trying to have a $K$ dispersion less than 2.5 mag, to avoid large differences 
 in the integration times and the spectral signal-to-noise ratios (SNR). 
 
 We included in the masks mainly stars with reddening-free parameters $Q_{IR}$ between $-0.2$ and 0.2 (characteristic of OB-type 
 stars, \citealt{ramirezalegria12}) and $K$ less than 13 mag. Some stars which do not fulfill the $Q_{IR}$ criteria, but do not 
 overlap with the OB-type candidate slits during mask design, were also included in both masks A and B (stars number 
 2, 7, 11, 20, and 21). The $Q_{IR}$ value for these five objects ranges between -0.27 and 0.27.
 
  The mask design also considered the spectral range derived from the slit position. For slits located in the positive half of the detector 
 (from the center to the right), we obtain spectra from 1.55 to 1.85 $\mu m$ in the $H-$band and from 2.06 to 2.40 $\mu
 m$ in the $K-$band. These spectral ranges include the \ion{He}{I} 1.70 $\mu m$, \ion{He}{I} 2.11 $\mu m$,
 \ion{He}{II} 2.19 $\mu m$, and \ion{He}{II} 1.69 $\mu m$ lines, which are required for early-type stellar spectral 
 identification and classification. 
  
  Both masks were observed using the $H$ and $K$ pseudogrism. The seeing values during observations, measured 
  from the telluric lines in the spectra, were between 0.75 and 1.40 arcsec. Individual slits vary between 8.5 and 10 arcsec 
  long and are 0.85 arcsec wide.
  
  For long-slit spectroscopy we selected bright and red (magnitude $K_S < 9$ and colour $(J-K_S) > 2.0$) objects as red 
  supergiant candidates. For these candidates, we do not expect a reddening-free parameter $Q_{IR}\sim0$, therefore for 
  long-slit spectroscopy candidates we relax this restriction. We observed a total of seven stars (objects number 24, 
  25, 26, 27, 28, 29, and 30) separated in four slits of 0.75 arcsec wide. For both long-slit and MOS observations, the resolution 
  was $\lambda/\Delta\lambda = 1500-1700$.
   
  To remove the sky contribution in the data reduction, we observed using an ABBA patern (the star is located in positions 
 A and B in the slit, the positions then being sequentially changed). Flat-fielding, spectral tracing, sky subtraction, coaddition, 
 and extraction were applied using {\sc iraf}\footnote{{\sc iraf} is distributed by the National Optical Astronomy 
 Observatories, which are operated by the Association of Universities for Research in Astronomy, Inc., under 
 cooperative agreement with the National Science Foundation.} for the long-slit spectra and {\sc lirisdr}\footnote
 {http://www.iac.es/project/LIRIS}, a package developed specifically for LIRIS data, for the MOS spectra.
  
   Combining the individual spectra, we discarded cosmic rays and hot pixels that might mimic spectral lines. 
 For wavelength calibration, argon and xenon lamps were observed, both lamps (continuum-subtracted) being used to 
 calibrate the $K$-band spectra and the argon lamp only for the $H$-band spectra. The telluric subtraction was carried
  out using {\sc molecfit} \citep{kausch15,smette15}, a software that corrects the atmospheric absorption lines by fitting 
  a synthetic transmission spectrum. The software models the synthetic spectrum by estimating the column density of 
  atmospheric molecules (water, carbon dioxide, methane, nitrous oxide, and ozone) from the physical parameters during 
  the observing night (external and internal temperature, relativity humidity, observatory coordinates and altitude, wind speed, 
  atmospheric pressure). 

To quantify the spectra quality, we measured the signal-to-noise ratio (SNR) per resolution element using the {\sc iraf} task {\sc slot}. 
We used as spectrum continuum the spectral ranges 1.611-1.641 $\mu m$, 1.650-1.675 $\mu m$, 2.060-2.105
$\mu m$, and 2.190-2.247 $\mu m$, for early-type stars, and 1.675-1.711 $\mu m$ and 2.075-2.2080 $\mu m$, for late-type 
stars. Most of our spectra have SNR greater than 70 (necessary to detect weak absorption features for $R\sim1000$ or larger,
 \citealt{hanson96}), and the final values are included in Table \ref{data_stars}.

\section{Spectral classification}\label{spec}

\begin{figure*}
\centering
\includegraphics[width=7in,angle=0]{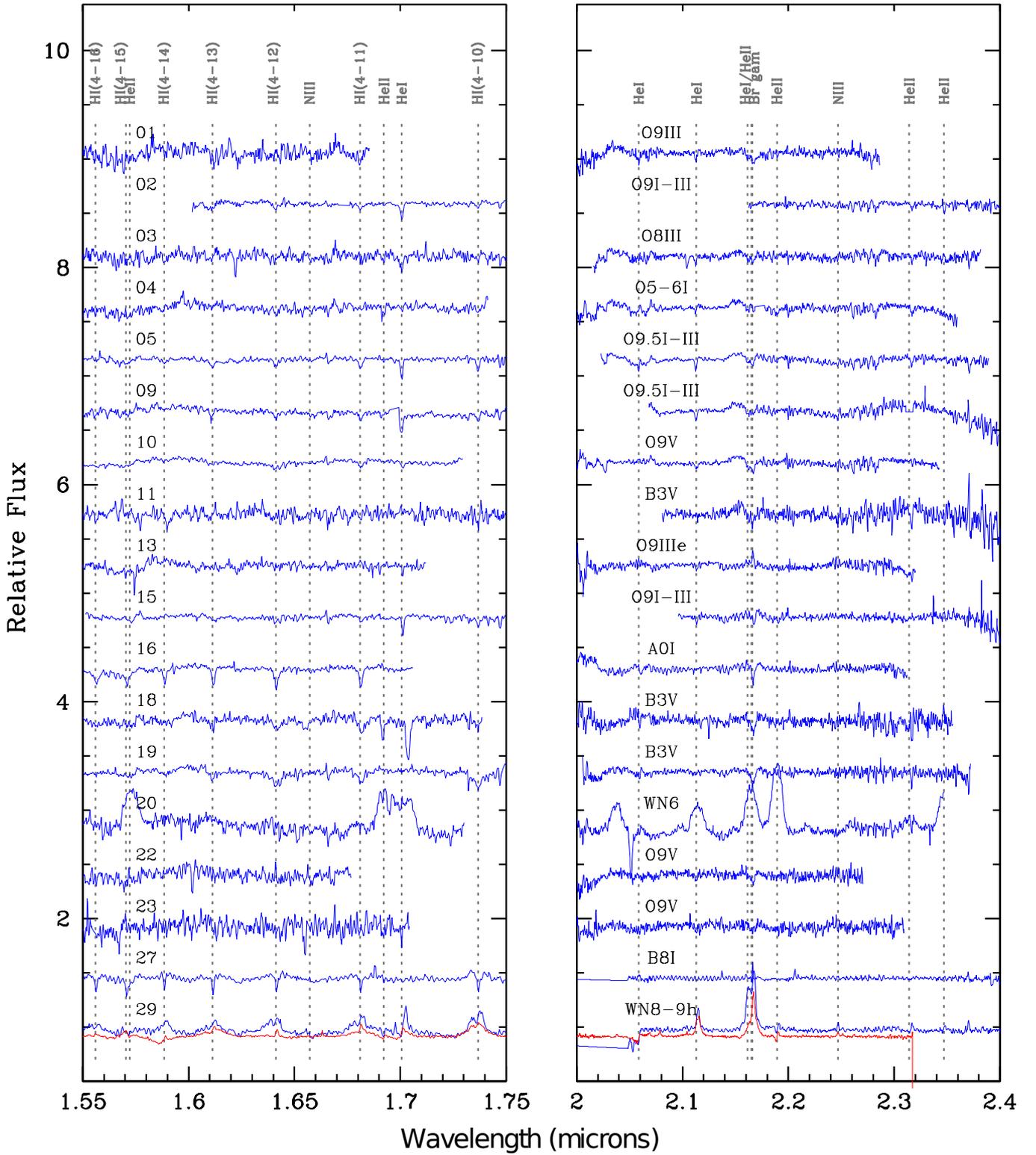} 
            \caption{Individual $H$-band (left) and $K$-band spectra (right) for massive stars. Labels indicate the object's ID
            ($H$-band spectra) and the spectral type ($K$-band spectra). This set includes early-type stars with luminosity classes I, 
            III, and V, and two Wolf-Rayets. The spectral features used for the spectral classifications are labelled in grey. The bottom 
            red spectrum plotted over spectrum29 corresponds to WR62-2, a WN8-9h object \citep{chene15} shown for visual comparison with the
            new Wolf-Rayet star WR122-16. Red supergiant spectrum is included in Figure \ref{m06_late}.}
       \label{m06_early}
\end{figure*}
 
\begin{figure*}
\centering
\includegraphics[width=7in,angle=0]{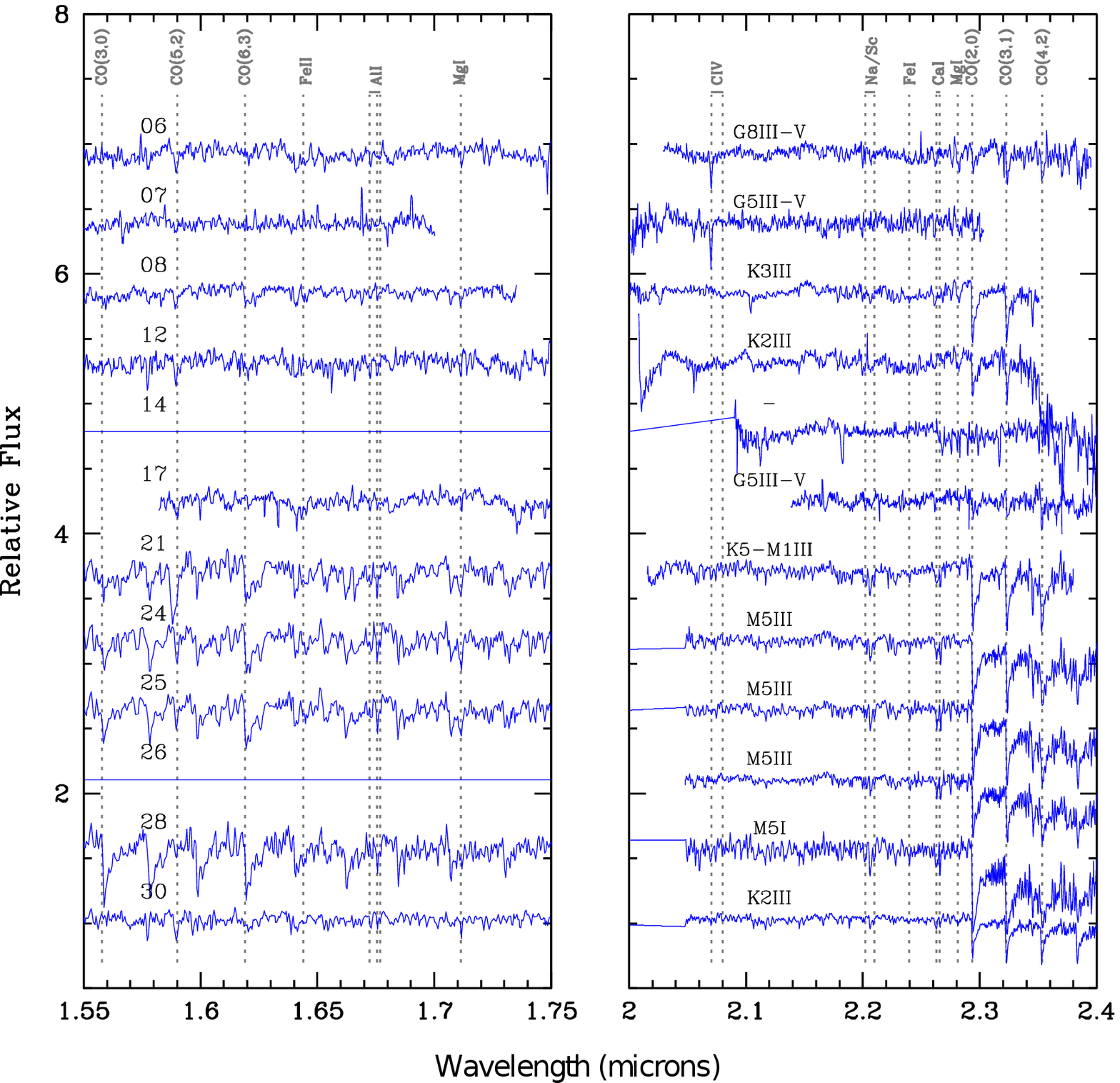} 
            \caption{Individual $H$-band (left) and $K$-band spectra (right) for late-type stars. In this figure
            we include objects with luminosity class III or V, the red supergiant star 28 and the unclassified object
            14. The spectral features used for the spectral classifications are labelled in grey. Labels in each spectrum
            also indicates the individual ID ($H$-band spectra) and the spectral type ($K$-band spectra).}
       \label{m06_late}
\end{figure*}

 We classify our spectra following the same procedure than for Masgomas-1 and Masgomas-4 \citep{ramirezalegria12,ramirezalegria14b}. 
 Using spectral libraries with similar resolution than LIRIS spectra, we identify spectral lines and compare the shapes and depths. For 
 early-type stars, we used the \citet{hanson96}, \citet{meyer98}, \citet{hanson98}, and \citet{hanson05} (degraded to our 
 spectral resolution) atlases. For later spectral types, we used \citet{wallacehinkle97} and the IRTF spectral library \citep{rayner03,cushing04}. 

The near-infrared early-type spectra show mainly the hydrogen Brackett series and helium (He\,I and He\,II) lines. The extension 
of the Brackett series is useful to set the upper limit (earliest spectral type) for the classification. The latest the spectral type, the most extended
will appear the Brackett series to the blue part of the spectra. For example, the H\,I (4-12) line at 1.64 $\mu m$ is barely noticed for stars earlier
than O5. The shape of the Brackett series helps to distinguish between luminosity class I (very narrow and deep lines) and III-V (broader and 
shallower lines).

The separation between luminosity classes III and V may be more challenging, but the shape and presence of He\,I and He\,II 
lines can help to the classification, for spectra with SNR$\sim$70 or greater. The presence of He\,I and He\,II lines are also good indicators for
the spectral type. For dwarf stars the presence of  He\,II at 1.69 $\mu m$ and/or 2.19 $\mu m$ lines indicates a spectral type not later than O7-8. 
In the case of the He\,I 1.70 $\mu m$, 2.06 $\mu m$, and 2.11 $\mu m$, the detection in the spectra sets a lower limit for the spectral type at
 B2-3. At our resolution and SNR, this set of lines allows a spectral classification of 2 subtypes, a similar uncertainty 
than reported by \citet{hanson10} and \citet{negueruela10}, with comparable datasets. For some of our stars, the luminosity classification is not 
possible, and we prefer to discard them from the analysis.

For late-type stars, the luminosity classification is based on the detection and equivalent width of CO($\nu$=3-6) 1.62 $\mu m$, 
CO($\nu$=0-2) 2.29 $\mu m$, and CO($\nu$=1-3) 2.32 $\mu m$. All spectra show the CO-band at 1.62 $\mu m$, discarding
the dwarf luminosity class, and only the spectra whose equivalent width between 2.294 and 2.304 $\mu m$ univocally indicates  
a supergiant luminosity class from the relation by \citet{davies07}, were classified as red supergiant. 

Most of our spectra show the features from OB-type stars, but luminosity classes range from dwarf to supergiants. The presence and shape 
of the Brackett series and the He\,I lines in absorption indicate that the earliest dwarf stars have a spectral type of O9. In the group of dwarfs 
objects, we classify stars number 10, 22, and 23 (classified as O9\,V), and stars number 11, 18 \& 19 (classified as B3\,V). For the first 
group the He\,I lines at 1.70 and 2.11 $\mu m$ and the Brackett series are clear in both $H$ and $K$-bands. The He\,I lines are clear but not as 
 deep as expected for the Brackett series in luminosity class III or I stars. The spectra are similar to the O9\,V HD93521 spectrum (standard 
 star, observed by our group using LIRIS). The spectra from the second group, classified as B3\,V, have a clear Brackett series 
 extending until the Br$\gamma$ H\,I (4-15) line. The He\,I 2.11 $\mu m$ is weak and fit the B3\,V HR\,5191 spectrum from \citet{meyer98}.
      
  Stars number 1, 3, and 13 are classified as early-type giants (O8-9\,III). The spectral features have similar depth and shape than observed 
  in stars HD\,37043 (O9\,III; \citealt{hanson96}), HD\,36861 (O8\,III; \citealt{hanson96}), and HR\,1899 (O9\,III; \citealt{meyer98}). For 
  star number 1, the He\,I 2.06-2.11 $\mu m$ lines are similar than observed in the O9\,III star HD\,37043 \citep{hanson96}. For star number
  3, we observe the depth of He\,I 1.70 $\mu m$ at a similar level than giant stars and the Brackett series shallower than observed in  
  O8-9\,III stars \citep{meyer98,hanson05}. In the case of star 13 the classification is mostly based on its broad and shallow Brackett series, 
  observing a  H\,I (4-7) peculiar emission line, plus its He\,I 1.70 $\mu m$ line, which is too deep to be considered a dwarf star.

 Stars 2, 5, 9, and 15 show spectral features observed in supergiants and giants stars. For these stars we could only determine the spectral type
  with a luminosity class between I and III. $K$-band spectrum of star 2 does not show distinctly spectral lines but in the $H$-band spectrum we observe
  a strong He\,I 1.70 $\mu m$ absorption line, more profound than the Brackett series. This characteristic is not seen in dwarf stars. The He\,I line
  is deeper than observed in the O9\, I HD30614 spectrum \citep{hanson98} but the Brackett series is too shallow to fit a supergiant luminosity class.
  Spectra of stars 5, 9, and 15 also present the He\,I 1.70-2.11 $\mu m$ line in absorption, deeper than the line observed in star 2, and the 
  Brackett series is observed until H\,I (4-15). The shape and depth of the He\,I is compatible with O9\,I stars \citep{hanson05}, but the depth of the
  Brackett series is in agreement with a giant luminosity class. The shape of the lines does not allow to determine a luminosity class for these objects.
	
  Star number 4 has the only spectrum observed with multi-object mask classified as an early supergiant. Its $H$-band spectrum does not 
  show the Brackett series, but we observe the He\,II 1.69 $\mu m$ line, deeper than the He\,I 1.70 and 2.11 $\mu m$ lines. The He\,II/He\,I 1.69/1.70 
  ratio indicates an O5-6\,I spectral type (for example Cyg OB2 8C; \citealt{hanson05}. The C\,IV 2.08 $\mu m$ in emission also supports an O5-6\,I spectral 
  type for this object.
  
The spectrum of star number 16 shows a narrow Brackett series both in $H$ and $K$-band, which is characteristic of A-supergiants. We classify
this object as A0\,I, after comparing with star A0\,I \object{HR 3975}.
      
 Star number 20 shows broad emission lines and corresponds to the Wolf-Rayet \object{WR122-11}. The broad He\,I 1.70, 2.06, 2.11 and 
 He\,II 1.57, 1.69, 2.03, 2.16, 2.19, 2.35 lines in emission dominates both $H$ and $K$-band spectra, and the Brackett series appears 
 weakly in emission. Our $K$-band spectrum is similar to the WR134 spectrum from \citet{figer97}, and we confirm the WN6 classification for 
 this object, given by \citet{faherty14}.
 
 The long slit spectra include very bright stars, RSG candidates and bright OB-type candidates who could not be included in the 
 mask. From this group of 7 spectra we classify 4 as massive evolved objects (stars 25, 27, 28, and 29).
 
  Star 25 shows a deeper CO band at 1.62 $\mu m$, compared with stars 24 and 26. The Al\,II, Ca\,I, and Mg\,I lines fit a supergiant 
 (we used star M1\,I-HD14404 from IRTF library as example) and a giant luminosity class (for example, star M5\,III HD175865). For this star, 
 the equivalent width of the ${}^{12}$CO (2,0) band in the region between 2.294 and 2.304 ${\mu m}$ is $\mathrm{{EW}_{25}}=23.29 \AA$.
 In the relation given by \citet{davies07}, this star could be classified as mid-M giant or early-M supergiant. Thus we can not 
 determine the luminosity class for this star and we discard it from the analysis.

 The spectrum of star 27 (the blue supergiant/LBV candidate \object{IRAS 18576+0341} \citealt{ueta01,pasquali02}) shows the Brackett series 
 with a central narrow absorption profile mixed with a broader emission profile. The depth of the absorption line component is similar to that
 observed in B8 supergiants (for example, HR1713), but the emission component could affect the natural depth of the series. We also observed 
 in emission two Na\,I lines at 2.206 and 2.209 $\mu m$, and one Mg\,II at 2.137 $\mu m$. 
 
 When we compare our spectrum with the one published by \citet{clark09}, we observe weaker emission features, and the previously stronger
  emission Brackett $\gamma$ H\,I (4-7) line now mixed with an absorption central feature (not seen in the spectra presented by \citealt{clark03,clark09}). 
  This mixed profile in the  Brackett $\gamma$ line is also observed in the  $H$-band Brackett lines. We classify this star as B8\,I, in agreement 
  with \citet{clark03} and because of the emission contaminating the photometry, we prefer to discard this star  from the distance and extinction 
  estimations, assuming the distance of 10 kpc estimated by \citet{ueta01}.
  
  Star 28 has clear and deep CO-bands in both the $H-$ and $K-$band spectra. The depth of both CO-bands implies a supergiant luminosity
class. Comparison with IRTF spectra indicates that star 28 is later than M5\,I. The equivalent width of the ${}^{12}$CO (2,0) band
is $\mathrm{{EW}_{28}}=34.66 \AA$, in the region between 2.294 and 2.304 ${\mu m}$. The EW and the shape of the spectrum 
point to a spectral type of M5\,I for this star.
 
 Star 29 spectrum has a broad Brackett series in emission. The He\,I (at 1.70, 2.11, and 2.16 $\mu m$) and N\,III (at 2.11 and 2.25 $\mu m$) 
lines are evident in emission. The He\,II can be observed in absorption clearly at 1.70 and 2.19 $\mu m$ (this last one present a P-Cygni profile). 
The spectrum is similar to a WN object (O4-6If+/WN9 VVVCL\,73-2 by \citealt{chene13}, WN8-9h WR\,62-2 by \citealt{chene15} or F2 and F7
by \citealt{martins08}), without a narrow C\,IV emission line detected in our spectrum. We classify star 29 as WN8-9h and name it as 
\object{WR122-16}, following the nomenclature by the ``Galactic Wolf Rayet Catalogue'' (version 1.19, P. Crowther, private communication).
   
The contaminants in this group are characterized by CO-bands in both the $H-$ and $K-$band spectra. But the depth and shape of this CO-bands
and the observed metallic lines discard them as supergiant objects. The shape of the CO-band at 1.619 $\mu m$ of star 24 resembles an M-late 
giant spectrum (for example, the M5\,III HD175865 from IRTF library). In the case of star 26, we only have $K$-band spectrum. The shape and 
depth of its CO-band and the Ca\,I lines indicate an M-late giant type (same as star 24). Star 30 spectrum also shows ${}^{12}$CO $\Delta \nu=3$ 
and ${}^{12}$CO $\Delta \nu=2$ bands, but with depths similar to a giant luminosity class. We assigned to this star the spectral type K2\,III.
  
\section{Analysis}\label{analysis}

From the 30 spectra, we estimated the individual distances, extinction and radial velocities (RVs). Using together 
the photometrically derived distances and the spectroscopically derived RVs, we characterize the massive star population of the cluster
candidate Masgomas-6, discovering the presence in the candidate area of two different groups of massive stars, separated by 4 kpc.

\subsection{Individual distances}

 To estimate the individual distances for the stars with spectral classification we compared the apparent magnitude with the intrinsic 
 magnitude corresponding to their spectral type, assuming the \citet{rieke89} extinction law, with $R=3.09$ \citep{rieke85}.
 This selection of extinction law was the same used for the discovery of the cluster candidate. The intrinsic magnitudes and colours for 
O-type stars, for all luminosity classes, are from \citet{martins05}. For stars later than O9.5\,V, we used the intrinsic magnitudes and 
colours from \citet{cox00}. The spectral type uncertainty dominates the distance errors, and we estimated it by deriving the 
individual distance for the same star assuming $\pm2$ spectral subtypes, except for the blue supergiants (BSG) stars number 2, 4, 5, and 9.

For these objects intrinsic magnitudes by \citet{martins05} have practically no variation (meaning extinction and distance estimates almost
constant in the $\pm2$ spectral subtypes range). Using the apparent magnitudes from the BSG-population in Arches \citep{martins08} and 
Mercer 30 \citep{delafuente16}, we estimated a dispersion of $\sim0.6$ mag in $K$-band per 2 spectral subtypes. We used this dispersion
in magnitude to estimate the uncertainties in extinction and distance estimates for the BSG in our sample.

 For the Wolf-Rayet objects, we used the absolute magnitude calibration by \citet{rosslowe15}. From a sample of 126 Wolf-Rayet with known
 distances, the authors derive near-infrared $JHK_S$ absolute magnitudes for nitrogen, carbon and oxygen-type WR. 
 
  In Table \ref{data_stars} we present the individual extinction and distance determinations for the spectroscopically observed stars. We distinguish
 two groups of distances for the massive stars: one nearby group (placed between 4 and 5 kpc) and one distant group
 (with distances greater than 6.5 kpc). Because both populations lay in the same line-of-sight, to decontaminate the colour-magnitude 
 diagram using statistical field-star decontamination is also not possible, and to derive physical parameters for the clusters or associations, (such as total mass by IMF integration or
 age determination via isochrone fitting) from the cleaned photometric diagrams is not possible. Using only the decontaminated colour-magnitude 
 and colour-colour diagrams (Figures \ref{m06_cmd} and \ref{m06_ccd}) is not possible to distinguish two or more stellar populations at different 
 distances, based on differences in the extinction law (across the line-of-sight or a product of differential reddening in any of the clusters or associations). 
 
 \subsection{Radial velocities}
 
We used the {\sc iraf} task {\sc rvidlines} to measure radial velocities from the observed spectra. The task compares the wavelength shift in spectral
absorption lines relative to specified rest wavelengths (Table \ref{wavelength} includes the rest wavelength used for both early and late-type objects).
After measuring a series of wavelength shifts, the task estimates their average and the radial velocity. This task is more suitable to 
determine radial velocity for low signal-to-noise or few lines spectra, compared with task based on template cross-correlation (such as {\sc fxcor}).
The line centering includes both a standard centering algorithm, similar to the {\sc identify} task, or a Gaussian fitting and deblending centering algorithm. 
Using the task {\sc rvcorrect} from the same {\sc iraf rv} package, we corrected the radial velocities to the local standard of rest, considering the observation 
date and time for each spectrum.

Because of the few number of lines identified in each spectrum, the errors of the LSR radial velocities are large and strongly dominated by the error 
derived from the initial {\sc rvidlines} estimates. We also included quadratically the error associated with the {\sc rvcorrect} task, but they are a smaller 
contribution to the total error. The values of the corrected radial velocity and the associated error for each star are given in Table
\ref{data_stars}. We do not report radial velocities for stars number 14 (without spectral classification), 20 (WN6), 27 (B8\,I), and 29 (WN8-9h). 
In the first case, we could not observe spectral features in the stellar spectrum, and for stars 20, 27, and 29 we expect a wind dominated line-profile.
 
 \begin{table}
\caption{Spectral lines wavelength. References: (1) \citet{NIST_ASD}, (2) \citet{cox00}.}
\begin{center}
\begin{tabular}{ccc}
\toprule
Vacuum wavelength & Element  & Reference \\
$[\mu m]$                 &                &   \\
 \midrule
 \multicolumn{3}{l}{Early-type stars:} \\
 \midrule
1.556072 & H\,I (4-16)  & (1) \\
1.570497 & H\,I (4-15)  & (1)  \\
1.588490 & H\,I (4-14)  & (1)  \\
1.611373 & H\,I (4-13)  & (1)  \\
1.641169 & H\,I (4-12)  & (1)  \\
1.681113 & H\,I (4-11)  & (1)  \\ 
1.69230 & He\,II (7-12)  & (2)   \\
1.700711 & He\,I $4d^3D-3p^3P^0$ & (1)   \\
1.736687 & H\,I (4-10)  & (1)   \\
2.058690 & He\,I $2p^1P^0-2s^1S$ & (1)   \\
2.112583 & He\,I $4s^3S-3p^3P^0$ & (1)   \\
2.166121 & H\,I (4-7)  & (1)   \\
2.189110 & He\,II (7-10)  & (2)   \\
\midrule
\multicolumn{3}{l}{Late-type stars:} \\
\midrule
1.61890  & CO($\nu$=3-6) & (2) \\
1.71133  & Mg\,I $4sS_1-4pP^{0}_{1}$  & (1) \\
2.20624  & Na\,I $4p^2P^0_{3/2}-4s^2S_{1/2}$  & (2)   \\
2.20897  & Na\,I $4p^2P^0_{1/2}-4s^2S_{1/2}$  & (2)   \\
2.26311  & Ca\,I $4f^3F^0_{3}-4d^3D_{2}$  & (2)   \\
2.26573  & Ca\,I $4f^3F^0_{4}-4d^3D_{3}$  & (2)   \\
2.29353  & CO($\nu$=0-2) & (2) \\
2.32265  & CO($\nu$=1-3) & (2) \\
\bottomrule
\end{tabular}
\end{center}
\label{wavelength}
\end{table}

 \subsection{Masgomas-6a and Masgomas-6b}
 
\begin{figure}
\centering
\includegraphics[width=9.8cm,angle=0]{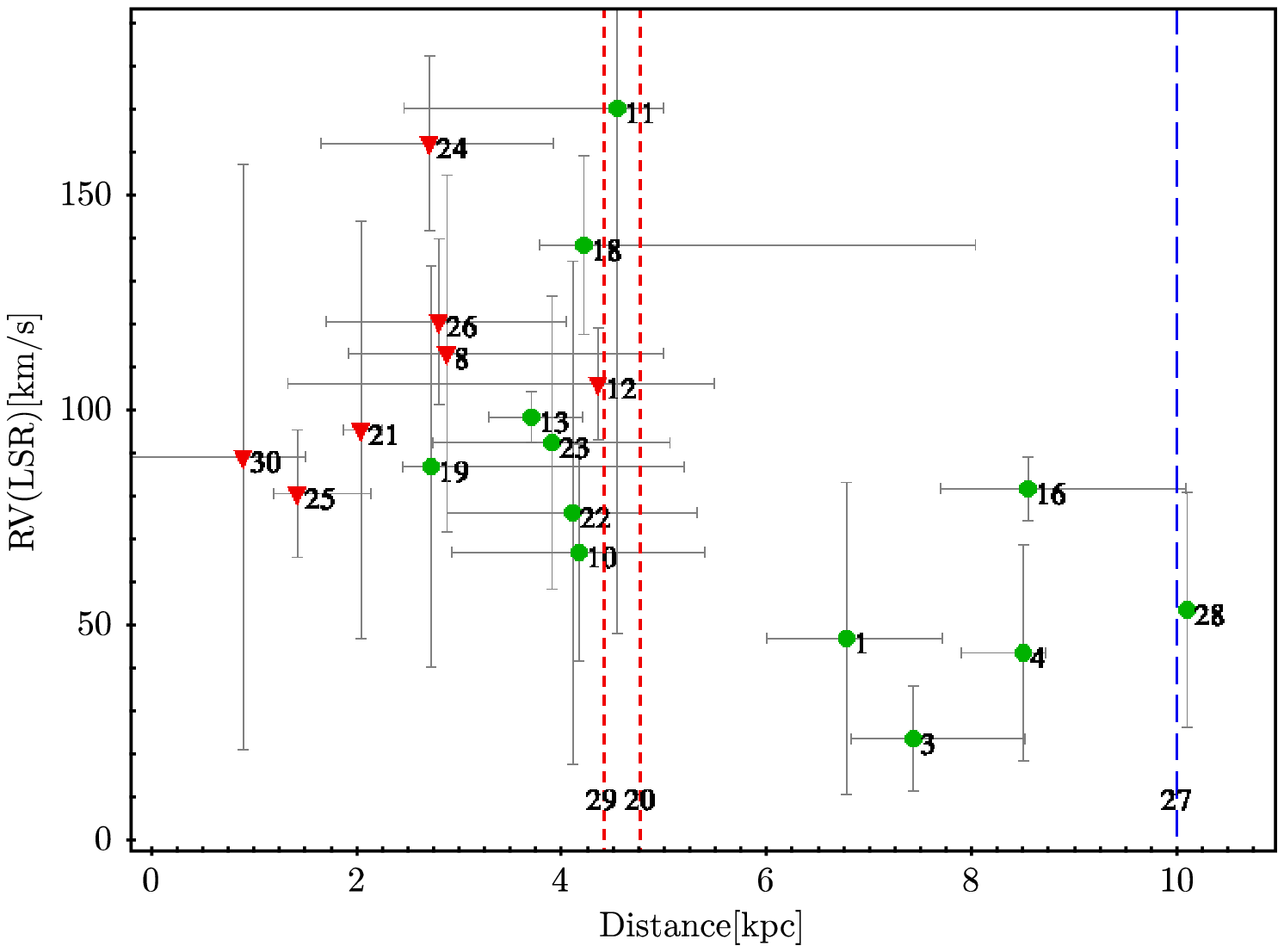} 
            \caption{Individual distance and radial velocity estimates for stars with spectroscopic observations. Objects are
            identified using the ID number from Table \ref{data_stars}, and shown using green circles (massive stars) and 
            red triangles (late-type giants). The vertical lines mark the distance estimates for the Wolf-Rayet stars (red line, objects 
            number 20 and 29) and the LBV (blue line, object 27). Only stars with assigned luminosity class are included in the figure.}
       \label{rv-distance-extinction}
\end{figure}

 Using together the individual distances, extinctions and radial velocity estimates, we can improve the identification of the massive star populations at
 different distances. In Figure \ref{rv-distance-extinction}, we see that for distant objects (i.e., d $>$ 6kpc, with the exception of star 16) the mean 
 radial velocity is below 50 km/s (no distant stars with large radial velocity), and for close objects the radial velocities range between 50 and 180 km/s. Comparing
 with the radial velocity measures for masers by \citet{reid14}, we see that in the galactic direction $l=37-38^{\circ}$, the expected radial velocity
 for distant stars is also less than 50 km/s, and are almost exclusively members of the Perseus arm. In the same galactic longitude but for nearby 
 stars, we find a population mixed with the Scutum-Centaurus and the Sagittarius arms.
 
 Late-type stars (Table \ref{data_stars}; stars 6, 7, 8, 12, 17, 21, 24, 25, 26, \& 30) present radial velocities in the whole range as expected for stars
 closer than 4 kpc. These object should belong to the close part of the Sagittarius arm and also to the tangent section of the Scutum-Centaurus 
 (including the base of this arm and the near end of the Galactic bar). Due to our uncertainties in the distance determination, it is not possible to 
 separate the population for the close arms.
  
  In the case of nearby early-type stars, objects 10, 22, 23 (O9\,V stars), and 13 (O9\,IIIe) present distances and radial velocities compatible with 
  a single cluster or association. The mean distance (3.9$^{+0.4}_{-0.3}$ kpc), and radial velocity (96$\pm$6 km/s) place this population as part of the close 
  intersection of the Sagittarius arm. We refer to this group of massive stars as Masgomas-6a and may include the two Wolf-Rayet: the WN6 
  \object{WR122-11} (star 20 at 4.8 kpc) and the WN8-9h \object{WR122-16} (star 29 at 4.4 kpc), as part of its stellar population. Stars 11 and 18 (both B3\,V) 
  have a similar distance (mean distance of 4.4 kpc), but their larger radial velocities would place them in the Scutum-Centaurus arm. Star 19 (B3\,V) 
  exhibits a very low radial velocity, and because it is the limit of the massive star classification we prefer not to consider it as part of Masgomas-6a.
    
  The group of distant stars includes the blue (number 4), yellow (number 16) and red (number 28) supergiants, plus the 
  O-giants (stars number 1 and 3). All objects in this group (except for star 16-A0\,I,) show a radial velocity below 50 km/s and distance estimate greater
  than 6.0. We include the LBV candidate in this group (star 27, classified by us as B8\,I) based on the distance estimate by \citet{ueta01}. We do 
  not estimate either radial velocity or extinction for this object, but the previously estimated distance of 10 kpc for the LBV agrees with its membership 
  to the distant group (hereafter Masgomas-6b). The individual distances for Masgomas-6b indicates a single distance of 9.6$\pm$0.4 kpc, 
 mean extinction of 2.5 mag and radial velocity of 64$\pm$6 km/s. 
 
 Assuming that both Masgomas-6a and Masgomas-6b populations from a stellar cluster or association can be described using a Kroupa \citep{kroupa01} initial mass function (IMF), 
 we fitted it to the observed massive stars and integrated this function to derive a lower limit for the total mass of each cluster/association. For 
 Masgomas-6a, we have 6 objects with masses between 15 and 40 $M_{\odot}$, limits set by the O9\,V star \citep{martins05} and the WN8-9h 
 Wolf-Rayet \citep{ekstrom12}. 
 
  For a crude estimate of the total mass (lower limit) of Masgomas-6a, we fitted the Kroupa IMF to the middle
 point in the mass range (i.e., 27.5 $M_{odot}$), with the uncertainty derived from the spectral type determination of 2 subtypes (i.e., mass between
 18.8 and 42.5 $M_{\odot}$). After integrating the IMF from log $(M) = -1.0$ dex to 1.6 dex, we derived for Masgomas-6a a total mass (lower limit) 
 of $(9.0-1.3)\cdot10^3 {M}_{\odot}$. Following the same procedure, we found for Masgomas-6b a lower limit for its total 
 mass of $(10.5-1.5)\cdot10^3 {M}_{\odot}$. For both objects, the presence of evolved massive stars (the WN8-9h Wolf-Rayet in Masgomas-6a
 and the LBV-candidate in Masgomas-6b) sets a limit for the age of 5 Myr, similar to the age estimated by \citealt{chene13} for VVV\,CL009.
 
\section{Conclusions}

We present the spectroscopic confirmation of a massive star population around the first cluster candidate derived from an automatic search 
software based on AUTOPOP \citep{garcia09,garcia11}, and adapted to use 2MASS photometry. The spectroscopic observations of 30 stars from 
the candidate Masgomas-6 revealed two populations of massive stars in the same line-of-sight: Masgomas-6a and Masgomas-6b.

 The stellar population of both groups appears mixed in the colour-magnitude and colour-colour diagrams, and we could not observe differences in the
 extinction law to help a photometric distinction of both populations. Using the spectroscopic data, particularly the individual distance, extinction and
 radial velocity estimates, we could distinguish both groups and estimate their distances ($\sim3.9$ and $\sim9.6$ kpc) and radial velocities ($\sim96$ 
 and $\sim64$ km/s) for Masgomas-6a and Masgomas-6b, assuming that each one of them forms a physical entity.
 
 Masgomas-6a is part of the Scutum-Centaurus arm, close to the base of the arm and the near end of the Galactic bar. This 
 region of intense star formation contains six red supergiant clusters: RSGC1 \citep{figer06}, RSGC2 \citep{davies07}, RSGC3 \citep{clark09}, 
 Alicante\,8 \citep{negueruela10}, Alicante\,7 \citep{negueruela11}, and Alicante\,10 \citep{gonzalezfernandez12}, plus Masgomas-1, the first massive 
 stellar cluster found by our group \citep{ramirezalegria12}. The Galactic longitude of Masgomas-6a would place it on the outer edge of the arm, but 
 still very close by distance and direction to this very active star-forming region. With Masgomas-6a we also reported the discovery of a new Wolf-Rayet 
 object, WR122-16, and associated the previously known \object{WR122-11} as part of their massive population.
 
  The discovery of Masgomas-6b and its evolved-massive population confirms that the LBV candidate \object{IRAS 18576+0341}
  is not in isolation. The seven evolved massive objects are part of the Perseus arm and have a common radial velocity. The extension 
  and shape of both massive groups need to be confirmed with an extended spectroscopic study of the region. 
  
 To determine if Masgomas-6a and Masgomas-6b are clusters or associations is beyond of our available data. The accuracy of our RV 
 measures does not allow us to study the internal dispersion on each population. A more in-depth kinematic study with more extensive and precise 
 measures of individual radial velocities is required to understand whether the massive stars are members of a cluster or an 
 OB-association. The future data should allow extending the study to the less massive population associated to any of the objects reported in this work.

\begin{acknowledgements}

 We are very thankful to Amparo Marco and Ignacio Negueruela for very fruitful discussion and help; to Karla Pe\~na Ram\'irez for her 
 collaboration during the text edition, and to the anonymous referee for comments on the article. S.R.A. thanks to the support by the FONDECYT 
 Iniciaci\'on project No. 11171025 and the CONICYT + PAI ``Concurso Nacional Inserci\'on de Capital Humano Avanzado en la Academia 
 2017'' project PAI 79170089. J.B. acknowledges support by the Ministry for the Economy, Development, and Tourism's Programa Iniciativa 
 Cient\'{i}fica Milenio through grant IC120009, awarded to The Millennium Institute of Astrophysics (MAS). A.H. and K.R. acknowledge support 
 by the Spanish MINECO under projects AYA 2015-68012-C2-1 and SEV 2015-0548. MG acknowledges support from grants FIS2012-39162-C06-01, 
 ESP2013-47809-C3-1-R and ESP2015-65597-C4-1-R.

Based on observations made with the William Herschel Telescope (WHT). The WHT is operated on the island of La Palma by the Isaac Newton 
Group in the Spanish Observatorio del Roque de los Muchachos of the Instituto de Astrof\'isica de Canarias. This publication makes use of data 
products from the Two Micron All Sky Survey, which is a joint project of the University of Massachusetts and the Infrared Processing and Analysis 
Center/California Institute of Technology, funded by the National Aeronautics and Space Administration and the National Science Foundation. 
This research has made use of NASA's Astrophysics Data System. 
 
We estimated asymmetric errors using the Java applet by \citet{barlow04}, available at http://www.slac.stanford.edu/$\sim$barlow/java/statistics.html;
the statistical unpaired Student's t-test was completed using the ``Statistics to Use'' webpage (by Kirkman, T.W., 1996)

\end{acknowledgements}

\bibpunct{(}{)}{;}{a}{}{,} 
\bibliographystyle{aa} 
\bibliography{biblio}

\listofobjects

\begin{landscape}
\begin{table}
\caption{Spectroscopically observed stars. Identification number from the UKIDSS GPS catalogue, equatorial coordinates, 
near-infrared magnitudes ($J$, $H$, and $K_S$), spectral classification are given for all stars. For those stars with determined 
luminosity class, the estimated extinction and distance are also provided. Radial velocity and signal-to-noise ratio measured 
for the $H-$ and $K$-band spectra are included in the last two columns. ($^{a}$) Distance estimated by \citet{ueta01}. }
\begin{center}
\begin{tabular}{cccccccccccc}
\toprule
 ID & UKIDSS ID & RA (J2000) & Dec (J2000)  & $J$ & $H$ & $K$ & Spectral type & $A_{K}$ & Distance & Radial velocity & SNR ($H$ \& $K$)\\
      &                    & [deg] & [deg] & [mag] & [mag] & [mag]  &    &       [mag]        & [kpc]  & [km/s]  &  \\
 \midrule
 \multicolumn{10}{l}{Massive stars:} \\
 \midrule
\vspace{0.1cm}
1   & 439036534746 & 285.0448  & +03.7976  &  16.910$\pm$0.013  &  14.174$\pm$0.003    &  12.631$\pm$0.003  &  O9\,III  & 2.96$^{+0.01}_{-0.07}$ & 6.77$^{+0.92}_{-0.78}$  & 47$\pm$36 & 42--100  \\
\vspace{0.1cm}
2   & 439038129414 & 285.0139  & +03.7600  &  15.052$\pm$0.022  &  12.532$\pm$0.021    &  11.213$\pm$0.021  &  O9\,I-III   &  -- &  --   & 27$\pm$4 & 143--103  \\
\vspace{0.1cm}
3   & 439036534444 & 285.0427  & +03.7722  &  16.404$\pm$0.009  &  13.934$\pm$0.002    &  12.470$\pm$0.003  &  O8\,III  &  2.74$^{+0.01}_{-0.01}$ &  7.42$^{+1.09}_{-0.60}$  & 24$\pm$12 & 51--90  \\
\vspace{0.1cm}
4   & 439036535027 & 285.0476  & +03.7735  &  15.986$\pm$0.006  &  13.440$\pm$0.001    &  11.939$\pm$0.002  &  O5-6\,I  & 2.81$^{+0.13}_{-0.13}$ &  8.50$^{+2.04}_{-1.65}$  & 44$\pm$25 & 69--174  \\
\vspace{0.1cm}
5   & 439036534795 &  285.0459  & +03.7669  &  14.941$\pm$0.003  &  12.263$\pm$0.001&  10.698$\pm$0.030  &  O9.5\,I-III  &  --  &  --  & 21$\pm$6 & 154--148  \\
\vspace{0.1cm}
9   & 439036536349 &  285.0587  & +03.7423  &  15.569$\pm$0.005  &  12.653$\pm$0.001&  10.996$\pm$0.023  &  O9.5\,I-III  & -- &  --  & 31$\pm$13 & 111--125  \\
\vspace{0.1cm}
10  & 439036538142 &  285.0740  & +03.7523  &  12.764$\pm$0.021  &  11.714$\pm$0.027  &  11.073$\pm$0.001  &  O9\,V  &  1.25$^{+0.01}_{-0.05}$ &  4.17$^{+1.22}_{-1.24}$  & 67$\pm$25 & 153--149  \\
\vspace{0.25cm}
11  & 439036536804 &  285.0622  & +03.7357  &  14.349$\pm$0.002  &  13.673$\pm$0.002  &  13.127$\pm$0.004  &  B3\,V  &  0.89$^{+0.01}_{-0.04}$ &  4.53$^{+0.45}_{-2.08}$  & 170$\pm$122 & 48--63  \\
\vspace{0.1cm}
13  & 439038128600 &  284.9904  & +03.7543  &  15.395$\pm$0.004  &  12.755$\pm$0.001  &  11.234$\pm$0.001 &  O9\,IIIe &  2.88$^{+0.01}_{-0.01}$ & 3.69$^{+0.50}_{-0.43}$  & 98$\pm$6 & 67--112  \\
\vspace{0.1cm}
15  & 439036531697 &  285.0197  & +03.7478  &  12.506$\pm$0.046  &  11.414$\pm$0.041  &  10.739$\pm$0.029 &  O9\,I--III &  -- &  --  & 49$\pm$37 &  165--98 \\
\vspace{0.1cm}
16  & 439038130441 &  284.9973  & +03.7677  &  14.736$\pm$0.003  &  12.179$\pm$0.001& 10.752$\pm$0.034 &  A0\,I &  2.59$^{+0.03}_{-0.04}$ &  8.54$^{+1.54}_{-0.85}$  & 82$\pm$7 & 122--91 \\
\vspace{0.1cm}
18  & 439038131024 &  285.0100  & +03.7708  &  14.055$\pm$0.002  &  13.341$\pm$0.001  &  12.908$\pm$0.004 &  B3\,V &  0.82$^{+0.02}_{-0.02}$ &  4.21$^{+3.82}_{-0.44}$  & 138$\pm$21 & 80--79  \\
\vspace{0.1cm}
19  & 439036531005 &  285.0146  & +03.7719  &  13.428$\pm$0.031  &  12.599$\pm$0.001& 12.084$\pm$0.002 &  B3\,V &  0.95$^{+0.02}_{-0.02}$ &  2.72$^{+2.46}_{-0.29}$  & 87$\pm$47 & 145--151 \\
\vspace{0.1cm}
20  & 439036531671 &  285.0212  & +03.7909  &  16.463$\pm$0.009  &  13.641$\pm$0.002  &  11.825$\pm$0.002 &  WN\,6 & 2.83$^{+0.11}_{-0.08}$ & 4.77$^{+2.72}_{-1.17}$  & -- & 57--81 \\
\vspace{0.1cm}
22  & 439038230096 &  285.0128  & +03.8057  &  15.542$\pm$0.005  &  13.351$\pm$0.002  &  12.154$\pm$0.002  &  O9\,V &  2.37$^{+0.01}_{-0.05}$ &  4.10$^{+1.20}_{-1.22}$  & 76$\pm$59 & 52--88  \\
\vspace{0.25cm}
23  & 439038230150 &  285.0218  & +03.8059  &  16.348$\pm$0.009  &  13.827$\pm$0.002  &  12.409$\pm$0.002  &  O9\,V &  2.73$^{+0.04}_{-0.05}$ &  3.90$^{+1.15}_{-1.16}$  &  93$\pm$34 & 30--65\\
\vspace{0.1cm}
27  & 439036534825 &  285.0454  &  +03.76308  &  11.953$\pm$0.024  &   9.115$\pm$0.027  &   7.059$\pm$0.013 & B8\,I & -- & 10.00$^{+3.00}_{-3.00}$\,$(^{a})$  & -- & 92--112  \\
\vspace{0.1cm}
28  & 439036535555 &  285.0511  &  +03.74756  &  11.166$\pm$0.019  &   8.469$\pm$0.023  &  6.707$\pm$0.015 & M5\,I & 2.18$^{+0.01}_{-0.01}$ & 10.10$^{+0.40}_{-0.40}$ & 54$\pm$27 & 43--55  \\
\vspace{0.1cm}
29  & 439036535386 &  285.0460 &  +03.7930  &  13.792$\pm$0.001  & 10.807$\pm$0.025  &   9.123$\pm$0.022 & WN8-9h & 3.14$^{+0.17}_{-0.16}$ & 4.40$^{+1.30}_{-2.26}$  & -- & 95--94  \\
\midrule
\multicolumn{10}{l}{Field stars:}  \\
\midrule
\vspace{0.1cm}
6   & 439036534981 &  285.0474  & +03.7637  &  15.517$\pm$0.004  &  13.652$\pm$0.002    &  12.609$\pm$0.003  &  G8\,III--V  & 1.53$^{+0.02}_{-0.03}$ & 3.07$^{+0.44}_{-0.23}$  & 179$\pm$38 & 76--75  \\
\vspace{0.1cm}
7   & 439036537285 &  285.0660  & +03.7710  &  14.531$\pm$0.002  &  13.649$\pm$0.002    &  12.969$\pm$0.004  &  G5\,III--V  & 0.68$^{+0.02}_{-0.03}$ & 4.82$^{+0.42}_{-0.35}$  &  35$\pm$66 &  71--64 \\
\vspace{0.1cm}
8   & 439036537202 &  285.0658  & +03.7581  &  14.855$\pm$0.003  &  12.707$\pm$0.001&  11.357$\pm$0.000[9]  &  K3\,III  & 1.77$^{+0.09}_{-0.10}$ & 2.87$^{+1.10}_{-0.97}$  &  113$\pm$42 &  89--124 \\
\vspace{0.25cm}
12  & 439036538821 &  285.0792 & +03.7399  &  15.363$\pm$0.004  &  13.447$\pm$0.001  &  12.432$\pm$0.003  &  K2\,III  & 1.44$^{+0.07}_{-0.09}$ & 4.35$^{+1.14}_{-1.03}$  &  106$\pm$13 & 50--83  \\
\vspace{0.1cm}
14  & 439038127146 &  285.0133 & +03.7431  &  16.415$\pm$0.011  &  13.589$\pm$0.040  &  12.023$\pm$0.035 &  -- & -- & -- & -- &  72 \\
\vspace{0.1cm}
17  & 439036533079 &  285.0323 & +03.7516  &  14.734$\pm$0.003  &  13.372$\pm$0.001  &  12.553$\pm$0.003 &  G5\,III--V & 1.09$^{+0.02}_{-0.03}$ & 3.30$^{+0.28}_{-0.24}$  &  130$\pm$23 &  84--85 \\
\vspace{0.25cm}
21  & 439036532705 &  285.0296 & +03.7918  &  16.344$\pm$0.008  &  12.628$\pm$0.001& 10.601$\pm$0.022  &  K5--M1\,III  & 3.13$^{+0.02}_{-0.02}$ & 2.03$^{+0.20}_{-0.17}$  & 95$\pm$49 & 49--81  \\
\vspace{0.1cm}
24  & 439038128244 &  284.9972 & +03.7496  &   9.901$\pm$0.021  &    8.077$\pm$0.035  &   7.001$\pm$0.005 &  M5\,III & 1.10$^{+0.06}_{-0.02}$ & 2.71$^{+1.20}_{-1.06}$  & 162$\pm$20 & 39--68  \\
\vspace{0.1cm}
25  & 439036530552 &  285.0102  & +03.7507  &   9.601$\pm$0.019  &    7.482$\pm$0.040  &   6.447$\pm$0.013 &  M5\,III & 1.30$^{+0.11}_{-0.03}$ & 1.41$^{+0.72}_{-0.23}$  & 81$\pm$15 & 60--74  \\
\vspace{0.1cm}
26  & 439036532029 &  285.0217 & +03.7582  &   9.984$\pm$0.023  &    8.148$\pm$0.027  &   7.077$\pm$0.007 &  M5\,III & 1.10$^{+0.07}_{-0.02}$ & 2.80$^{+1.24}_{-1.10}$  &  121$\pm$19 & 87 \\
\vspace{0.1cm}
30  & 439036535381 &  285.0467 & +03.7886  &  10.055$\pm$0.023  &   8.893$\pm$0.035  &   8.231$\pm$0.021 & K2\,III & 0.71$^{+0.17}_{-0.16}$ & 0.88$^{+0.61}_{-1.17}$  & 89$\pm$68 & 70--140   \\
\bottomrule
\end{tabular}
\end{center}
\label{data_stars}
\end{table}
\end{landscape}

\end{document}